\documentclass[12pt,amssymb]{iopart}

\usepackage{amssymb}
\usepackage{graphicx}
\def\be{\begin{equation}}
\def\ee{\end{equation}}
\def\ep{\epsilon}
\def\iint{\int\!\!\!\!\!\!\!\int}
\def\ps{\tilde\psi}
\def\A{\tilde{A}}

\def\d{\rm d}
\def\i{\rm i}
\def\Name{}
\def\and{and\;}

\begin{document}
\title{Flux-induced Nernst effect in a superconducting loop}

\author{Jorge Berger}

\address{Department of Physics, Ort Braude College, 21982 Karmiel, Israel}
\ead{jorge.berger@braude.ac.il}

\begin{abstract}
When a superconducting ring encloses a magnetic flux that is not an integer multiple of half the quantum of flux, a voltage arises in the direction perpendicular to the temperature gradient. This effect is entirely due to thermal fluctuations. We study the dependence of this voltage on the temperature gradient, flux, position, average temperature, BCS coherence length, thermal coherence length, and the Kramer--Watts-Tobin parameter. The largest voltages were obtained for fluxes close to $0.3\Phi_0$, average temperatures slightly below the critical temperature, thermal coherence length of the order of the perimeter of the ring and BCS coherence length that is not negligible in comparison to the thermal coherence length. As a rough comparison between the flux-induced and the field-induced effects, we also considered a two dimensional sample.
\end{abstract}
\maketitle

\section{Introduction}
The Nernst–-Ettingshausen effect occurs when a temperature gradient and a magnetic field are present, and results in an electric field, perpedicular to the temperature gradient and to the magnetic field. 

In the case of simply connected superconductors, early experiments date to half a century ago in low temperature superconductors \cite{Ott,Munn,Hub}, and in high temperature superconductors the effect was detected \cite{Pal} soon after their discovery. Ullah and Dorsey evaluated the magnetothermoelectric coefficients \cite{Ul0,Ul1} within the framework of the time-dependent Ginzburg--Landau model (TDGL), ignoring cooperon contribution \cite{Varlamov,byn,Fi}, and obtained a giant value for the fluctuation Nernst effect above the superconducting transition for weak magnetic fields. Ussishkin {\it et al.} \cite{Ussi} noticed that the analysis of the Nernst effect should distinguish between transport and magnetization currents. Within the last decade there has been a renewed interest in magnetothermoelectric effects, leading to several theoretical \cite{Huse,Podo,Podo8,Ser,Baruch,AA,AS,ST,H9,var3,H10,Kot,var2,Bolech,Efrat} and experimental \cite{W0,W1,W2,Wang,Pourret,RA6,CC9,Daou,Hess,RA11} studies. Nernst signals have been observed in an extended region above the critical temperature in high-$T_c$ materials \cite{Wang} and in conventional superconductors \cite{Pourret}. The Nernst–-Ettingshausen effect is an active field of research also in graphene \cite{Check,Zuev,Ralf}.

In this paper we predict a qualitatively new effect, expected to appear in samples with ring topology: a thermoelectric voltage that is induced by the enclosed magnetic flux rather than by the magnetic field. We find that this voltage is largest in the direction perpendicular to the temperature gradient and is present when the magnetic flux is neither an integer nor a half-integer multiple of the quantum of flux.

A case with ring topology that has been studied experimentally \cite{HHG,Pb} and theoretically \cite{GZ,ec,Cz} is that of a bimetallic loop. A related effect, which has been reviewed in \cite{Galperin}, is the appearence of magnetic flux when a temperature gradient is present in a nonuniform loop. Another situation in which a flux dependent voltage appears in a nonuniform superconducting ring was reported in \cite{Nikul}.

In most of this paper we will examine the simplest possible configuration: a 1D ring of uniform material and uniform cross section.

The following section raises a naive argument of why a voltage is to be expected in the presence of flux and nonuniform temperature. In section \ref{MAP} we specify the considered system and the formalism for its description. In section \ref{ress} we evaluate the voltage in the TDGL limit and in section \ref{BTD} we extend our results beyond this limit. In section \ref{Dss} we speculate on the possibility of measuring the proposed effect. In the appendix we deal with 2D samples, where the field-induced and the flux-induced effects can both be present and can thus be compared.

\begin{figure}\center{
\scalebox{0.85}{\includegraphics{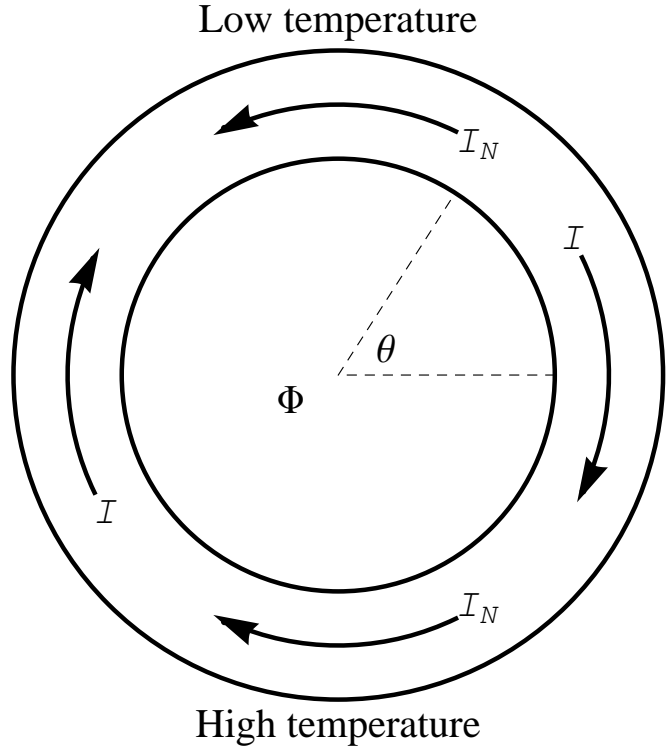}}}
\caption{\label{heuristic}Superconducting ring that encloses a magnetic flux $\Phi$. $I$ is the total current around the ring and $I_N$ is the normal current. }
\end{figure}

\section{Heuristic considerations\label{hc}}
We consider a superconducting ring with an average temperature close to its critical temperature $T_c$, that encloses a magnetic flux $\Phi$, as shown in fig.~\ref{heuristic}. For flux in the range $0<\Phi < 0.5\Phi_0$, where $\Phi_0=\hbar c/(2e)$ is the quantum of flux, a diamagnetic current $I$ flows around the ring \cite{Standord}. In the region where the temperature is higher than the average, superconductivity is weaker than the average and we could therefore conjecture that the supercurrent should be smaller than the average; as a consequence, a normal current $I_N$ would have to reinforce the supercurrent in order to reach the total current $I$. On the other hand, in the region of lower temperature, supercurrent should be large and the normal current would have to oppose it. The validity of this pair of conjectures will be discussed in section \ref{ress}. In order to maintain the normal current, a potential difference is required, higher in the region close to $\theta\approx 0$ in fig.~\ref{heuristic}, and lower in the region close to $\theta\approx\pi$.

\section{Model and procedure\label{MAP}}
We model the ring as one-dimensional, so that position in it is entirely determined by the angle $\theta$, as defined in fig.~\ref{heuristic}. 
We denote the average temperature by $(1+\epsilon )T_c$ and the temperature span by $2\delta T_c$, {\it i.e.} the highest and lowest temperatures along the ring are $T(\pm\pi /2)=(1+\epsilon\mp\delta)T_c$. Assuming linear dependence of the temperature on the position on the plane of the ring, the temperature around the ring is given by

\be
T(\theta )=(1+\epsilon -\delta\sin\theta )T_c \;.
\label{T}
\ee
Equation (\ref{T}) is justified if the heat transported by the ring is small in comparison to the heat transported through the substrate. If this is not the case, (\ref{T}) may be regarded as a simple interpolation.

Let us denote by $L$ the perimeter of the ring. Besides $L$, there are two characteristic lengths in this problem: the coherence length at zero temperature, $\xi (0)$, and a thermal length, $\xi_\beta=[w\Phi_0^2/(32\pi^3\kappa^2k_BT_c)]^{1/3}$, where $w$ is the cross section of the wire that makes the ring, $\kappa$ is the Ginzburg--Landau parameter and $k_B$ is the Boltzmann constant. From here we obtain a characteristic time $t_0=\xi_\beta^2/D$, where $D$ is the diffusion coefficient, and a characteristic voltage $V_0=\hbar /(2e t_0)$.

We choose a gauge in which the scalar electric potential vanishes. In this gauge, the voltage at position $\theta $ relative to position $\theta =0$ is given by $V(\theta )=[L/(2\pi c)]\int_0^\theta {\d}\theta '{\partial A}(\theta ')/\partial t$, where $A$ is the tangential component of the electromagnetic vector potential; the dependence of $A$ on the time $t$ is due to the Johnson noise. We discretize the problem by dividing the ring into $N$ segments of length $L/N$ and define the dimensionless quantities $\A=2\pi LA/N\Phi_0$ and
\be
s_0=0\;,\;\;\;s_k=s_{k-1}+\frac{1}{2}(\A_{k-1}+\A_k),
\label{sk}
\ee
where the subscript $k$ of a quantity denotes its value at $\theta =2k\pi/N$. In particular, $s_N=\sum_{k=0}^{N-1}\A_k=2\pi\Phi /\Phi_0$. With this notation, $V(\theta_k)=[\hbar /(2e)]\partial s_k/\partial t$. 

In order to follow the evolution of the $\A_k$'s, we use the Kramer--Watts-Tobin model \cite{KWT,Kopnin}, which successfully describes transport phenomena in superconductors, provided that there is local equilibrium. 
Addition of thermal fluctuations to the model and its discretized version have been described elsewhere \cite{JPCM}. Here we will just outline the main steps and adapt them to the present problem.

We make use of a gauge invariant order parameter in the form
\be
\ps_k=|\ps_k|\exp [{\i} (\chi_k+s_k)] \;,
\label{polar}
\ee
where $\chi_k$ is the argument of the single valued (not gauge invariant) order parameter. It follows from eq. (\ref{sk}) that $\ps$ obeys the periodicity condition $\ps_{k+N}=\exp (2\pi {\i}\Phi/\Phi_0)\ps_{k}$. We also note that in the 1D limit the induced flux is negligible, so that $\Phi $ is just the applied magnetic flux. The applied flux is kept constant, and we therefore have
\be
\sum_{k=0}^{N-1}\A_k(t)=\mbox{constant}\;.
\label{const}
\ee

Let $\tau$ be a period of time that is short compared to the relaxation times of the system. The evolution of $\A_k$ is given by \cite{Kopnin,JPCM}

\be
\A_k(t+\tau)=\A_k(t)-2.84(\tau /t_0)\left[{\rm Im}\left(\ps_{k-1}^*\ps_k+\ps_k^*\ps_{k+1}\right)+\tilde{I}\right]+\eta_{Ak}\;,
\label{Aps}
\ee
where the asterisk denotes complex conjugation, $\tilde{I}$ is a Lagrange multiplier that is adjusted at every step to obey eq.~(\ref{const}), and $\eta_{Ak}$ is a random variable with zero average, gaussian distribution, and variance
\be
\langle\eta_{Ak}^2\rangle =5.68\frac{DL\tau T(\theta )}{N\xi_\beta^3 T_c}\;.
\label{varianceA}
\ee

Denoting by $\Delta$ the change of a quantity between $t$ and $t+\tau$, the evolution of the order parameter can be expressed as $\Delta\ps_k=[\Delta |\ps_k|/|\ps_k|+{\i}(\Delta\chi_k+\Delta s_k)]\ps_k$. $\Delta s_k=(\Delta\A_0+\Delta\A_k)/2+\sum_{j=1}^{k-1}\Delta\A_j$ can be evaluated by means of eq.~(\ref{Aps}); the other changes are given by 
\be
\Delta |\ps_k|=h_k\Gamma \left(\frac{h_k^2k_B T(\theta )}{|\ps_k|}-\frac{\partial G}{\partial |\ps_k|}\right)\tau+\bar\eta_{|\ps |}
\label{Dabs}
\ee
and
\be
\Delta\chi_k=-\frac{\Gamma }{h_k|\ps_k|^2}\frac{\partial G}{\partial \chi_k}\tau+\eta_{\chi}\;.
\label{Dchi}
\ee
Here $\Gamma =ND/(2\xi_\beta Lk_BT_c)$, $h_k=(1+K|\ps_k|^2)^{-1/2}$, $K\approx 15Dk_BT_c\tau_{\rm ph}^2/(\hbar\xi_\beta^2)$, where $\tau_{\rm ph}$ is the electron-phonon inelastic scattering time, $\bar\eta_{|\ps |}$ and $\eta_{\chi}$ are random variables with zero average, gaussian distribution, and respective variances $2h_k\Gamma k_BT(\theta )\tau$ and $2\Gamma k_BT(\theta )\tau /(h_k|\ps_k|^2)$. $G$ is the free energy, which can be written as

\begin{eqnarray}\hspace{-1cm}
G=&\frac{Lk_BT_c}{N\xi_\beta }\sum_{k=0}^{N-1} \left\{ \frac{\xi_\beta^2(T(\theta )-T_c)}{\xi^2(0)T_c}|\ps_k|^2+\frac{1}{2}|\ps_k|^4+ \right. \nonumber\\
&\left. \frac{N^2\xi_\beta^2}{L^2}\left( 2|\ps_k|^2-2|\ps_k||\ps_{k+1}|\cos  [\chi_{k+1}-\chi_k+\frac{1}{2}(\A_k+\A_{k+1})]+\A_k\tilde{I}\right)\right\} \;.
\label{G}\hspace{0.7cm}
\end{eqnarray}

The material parameters may be estimated assuming a free electron gas and dirty limit \cite{Kopnin}: 
\be
\xi^2(0)\sim\frac{\pi\hbar ^2k_F\ell_e}{12mk_BT_c}\;,\;\; \kappa ^2\sim \frac{0.021mc^2}{n_e e^2\ell_e^2}\;,\;\; D\sim \frac{2\hbar k_F\ell_e}{3m}\;,
\label{estimates}
\ee
where $k_F$ is the Fermi wave number, $\ell_e$ is the mean free path, $m$ is the mass of an electron pair and $n_e$ is the electron density. With these estimates, $K\sim 10^{23}T_c^2\tau_{\rm ph}^2\,[{\rm K^{-2}sec^{-2}}]$.

For situations in which $K|\ps_k|^2\gg 1$, it is difficult to follow numerically the evolution of $|\ps_k|$ and $\chi_k$, because in this case $\chi_k$ evolves much faster than $|\ps_k|$. We therefore started by studying the limit $K\rightarrow 0$, and then investigated the influence of $K$. In the limit $K\rightarrow 0$ the Kramer--Watts-Tobin model reduces to TDGL, and eqs. (\ref{Dabs}) and (\ref{Dchi}) can be replaced with

\begin{eqnarray}\hspace{-2.5cm}
\ps_k(t+\tau)=&\left(\ps_k(t)-(\tau /t_0)\left\{\left[(\xi_\beta /\xi(0))^2(T(\theta )-T_c)/T_c +|\ps_k|^2\right]\ps_k\right.\right.\nonumber\\
&\left.\left. +(N\xi_\beta /L)^2(2\ps_k-\ps_{k-1}-\ps_{k+1})\right\}+\eta_{1k}+{\i}\eta_{2k}\right)\exp [{\i}(s_k(t+\tau)-s_k(t))]\;;
\label{evps}\hspace{1cm}
\end{eqnarray}
where $\eta_{1,2}$ are random variables with zero average, gaussian distribution, and variance
\be
\langle\eta_{1k}^2\rangle =\langle\eta_{2k}^2\rangle =\frac{ND\tau T(\theta )}{\xi_\beta LT_c}\;.
\label{varianceps}
\ee

Since $\Phi$ enters our equations only through the periodicity factor $\exp (2\pi {\i}\Phi/\Phi_0)$, $V(\theta ,\Phi +\Phi_0)=V(\theta ,\Phi )$. In addition, since changing the sign of $\Phi$ is equivalent to a 180$^\circ$ rotation of the ring about the ``vertical" axis in fig.~\ref{heuristic}, $V(\theta ,-\Phi )=V(\pi-\theta ,\Phi)-V(\pi ,\Phi)$. It is therefore sufficient to study the range $0\le\Phi\le\Phi_0/2$.

In our calculations we took $N=12$ and $\tau /t_0=5\times 10^{-4}$. In each run, the initial values of $\ps$ were random; $2\times 10^8$ steps were performed to enable relaxation to a ``typical" state, and then $V(\theta )$ was averaged during $4\times 10^9$ steps.

\section{Results in the TDGL limit\label{ress}}

\begin{figure}\center{
\scalebox{0.85}{\includegraphics{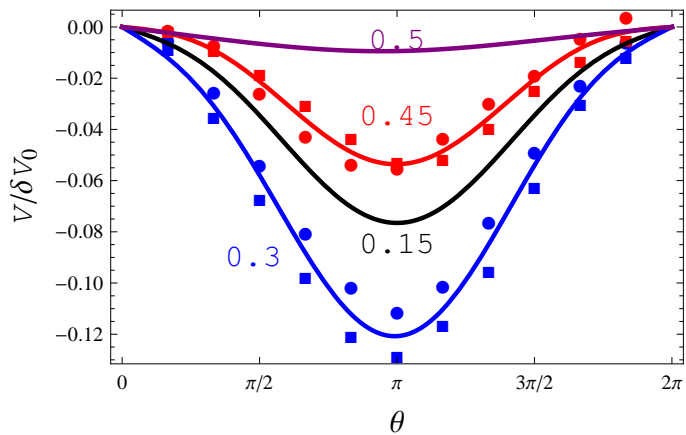}}}
\caption{\label{Voffi} Ratio of the voltage to the temperature difference, as a function of position, for several values of the flux $\Phi$. $\delta $ is half the difference between the maximum and the minimum temperature divided by $T_c$, and $V_0=\hbar D/2e\xi_\beta^2$ is the unit of voltage. The angle $\theta $ is defined in fig.~\ref{heuristic}. The ratio $\Phi/\Phi_0$ is marked next to each curve. The curves are smooth fits to the calculated values. Temperature and sample parameters: $\epsilon =-0.1$, $\delta=0.1$, $\xi_\beta/L=0.5$, $\xi_\beta/\xi (0)=3$.}
\end{figure}

Figure \ref{Voffi} is the central result of this paper. It shows the voltages $V(\theta )$ for a given ring and temperature profile, for several enclosed fluxes in the range $0<\Phi\le 0.5\Phi_0$. As expected from our heuristic considerations, these voltages are negative. The smooth curves are fits of the form $V(\theta )=\sum_{j=1}^3 v_j\sin(j\theta /2)$. For some of the curves we also show disks and squares, which represent the calculated values, for two different runs. The difference between the two runs provides an estimate for our statistical uncertainty. The statistical uncertainty may also be estimated from the value of $V(\pi )$ for $\Phi =0.5\Phi_0$, which has to vanish on symmetry grounds. For a given run, the values of $V(\theta )$ for different values of $\theta$ are highly correlated, due to the high ratio $\xi_\beta/L$ of the considered ring.

\begin{figure}\center{
\scalebox{0.85}{\includegraphics{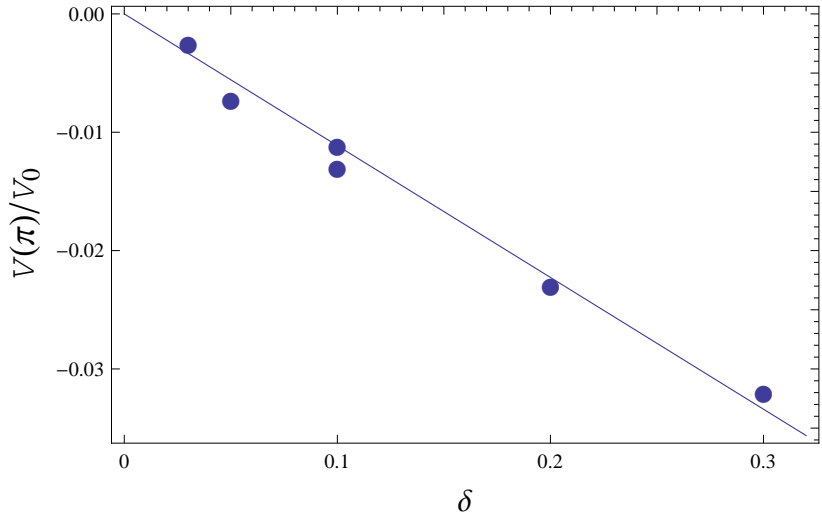}}}
\caption{\label{Vofdelta} Voltage at $\theta =\pi$, as a function of $\delta $, for $\Phi =0.3\Phi_0$. The average temperature and the sample parameters are as in fig.~\ref{Voffi}. $V(\pi )$ was obtained from smooth fits, as in fig.~\ref{Voffi}, and the straight line is a guide for the eye. }
\end{figure}

Figure \ref{Vofdelta} shows the voltage at a fixed position (the left extreme in fig.~\ref{heuristic}) and given flux, as a function of the temperature span. Our results are consistent with the proportionality $V\propto\delta$.

\begin{figure}\center{
\scalebox{0.85}{\includegraphics{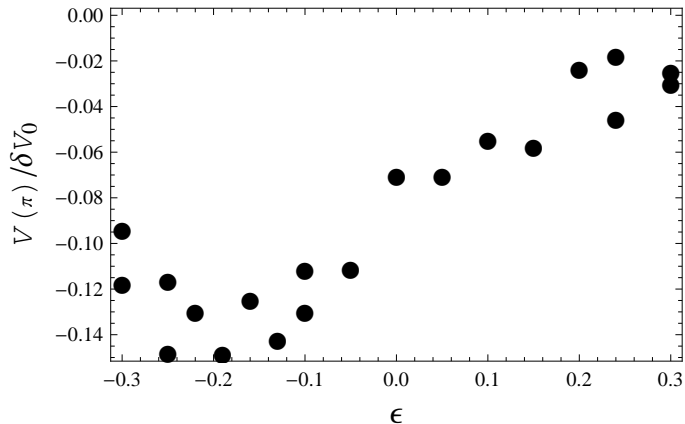}}}
\caption{\label{epsilon}Ratio of voltage to temperature difference, as a function of the average temperature. The enclosed flux is $0.3\Phi_0$ and the remaining parameters are as in fig.~\ref{Voffi}.}
\end{figure}

Figure \ref{epsilon} shows the voltage at a fixed position and given flux, as a function of the average temperature. Although the statistical uncertainty is large and TDGL is not reliable at the low temperatures involved, our results suggest that the Nernst effect is largest at $T\sim 0.8T_c$ and remains appreciable on a broad temperature range.

\begin{figure}\center{
\scalebox{0.85}{\includegraphics{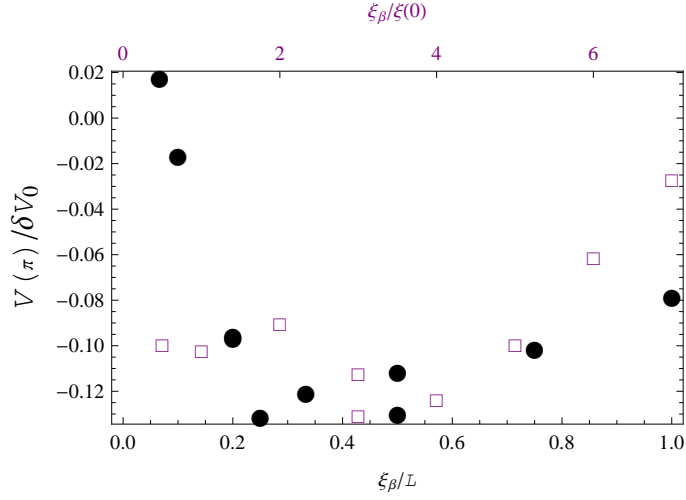}}}
\caption{\label{xx} Ratio of voltage to temperature difference, as a function of the ratios of $\xi_\beta $ to the perimeter (disks, lower scale) and to $\xi (0)$ (squares, upper scale). The enclosed flux is $0.3\Phi_0$ and the remaining parameters are as in fig.~\ref{Voffi}. The accuracy of our results for $\xi_\beta\lesssim 0.1L$ may be marred by the limited number of segments in the discretization; in the case $\xi_\beta = 0.2L$, doubling the number of segments did not change the result within the statistical significance.}
\end{figure}

Figure \ref{xx} shows the voltage dependence on the ratios $\xi_\beta /L$ and $\xi_\beta /\xi (0)$. The voltage is small for $\xi_\beta\lesssim 0.1L$, reaches a maximum size for $\xi_\beta\sim 0.3L$, and then decreases slowly as $\xi_\beta /L$ increases further. As a function of $\xi_\beta /\xi (0)$, $V(\pi )$ remains nearly constant for $\xi_\beta \lesssim 5\xi (0)$, but drops as $\xi_\beta /\xi (0)$ increases further.

\begin{figure}\center{
\scalebox{0.85}{\includegraphics{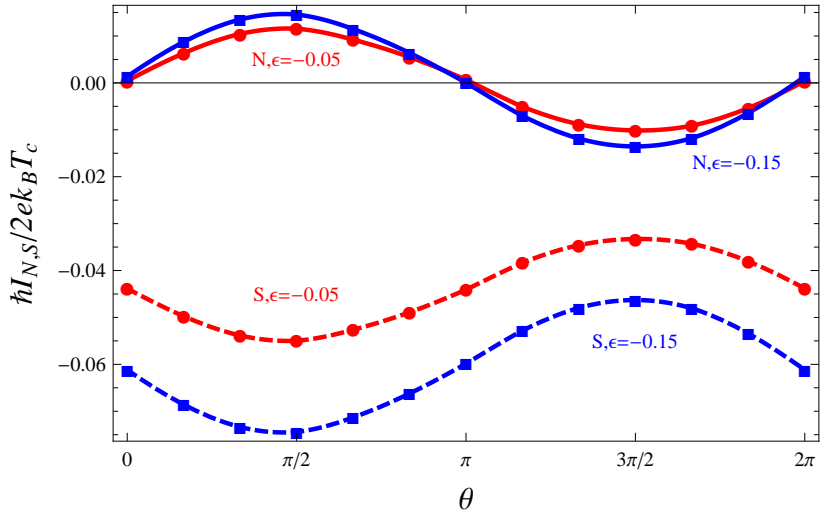}}}
\caption{\label{conjec}Normal current $I_N$ (solid lines) and supercurrent $I_S$ (dashed lines) as functions of position, for enclosed flux $\Phi =0.3\Phi _0$, for average temperatures $0.95T_c$ (red) and $0.85T_c$ (blue). $I_N$ and $I_S$ were averaged during a period of $5\times 10^5t_0$. Since $|I_S|$ is considerably larger than $|I_N|$, the curve $I_S(\theta )$ for $\epsilon =-0.05$ was raised by 0.35 units and the curve for $\epsilon =-0.15$ was raised by 0.8 units, in order to present $I_N$ and $I_S$ in the same graph. The markers were evaluated and the curves are interpolations. Other parameters: $\delta=0.2$, $\xi_\beta/L=0.5$, $\xi_\beta/\xi (0)=3$.}
\end{figure}

Let us now revise our conjecture in section \ref{hc}. Figure \ref{conjec} shows the position dependence of the superconducting and the normal currents for typical situations, and supports this conjecture. $|I_S|$ is indeed smaller in the region of higher temperature, and clockwise normal current is required in order to achieve the total current. We may also note that the normal current is not very sensitive to the average temperature.

It should be emphasized that the voltage that we obtain is entirely due to the inhomogeneity of thermal fluctuations. If in eqs.\ (\ref{varianceA}) and (\ref{varianceps}) we set the constant $T(0)$ instead of $T(\theta )$, whereas $T(\theta )$ is left unchanged in eq.~(\ref{evps}), the voltage becomes smaller than our uncertainty level. Moreover, there is no normal current in the absence of fluctuations. This result is expected from the second law of thermodynamics: in eq.~(\ref{evps}), position variation of $T$ is equivalent to position variation of $T_c$; therefore, if it were possible to obtain a potential difference for a particular temperature profile (without position dependent fluctuations), it would also be possible to obtain the same potential difference for a suitable profile of $T_c$ ({\it i.e.}, of material composition) at a uniform temperature. It is only through the fluctuations that the temperature gradient enters the problem in an essential way.

Accordingly, we must conclude that not every nonuniformity in the superconducting strength leads to nonuniformity of the supercurrent. Nonuniformity of thermal fluctuations does, whereas nonuniformity of material or cross section should not.

\begin{figure}\center{
\scalebox{0.85}{\includegraphics{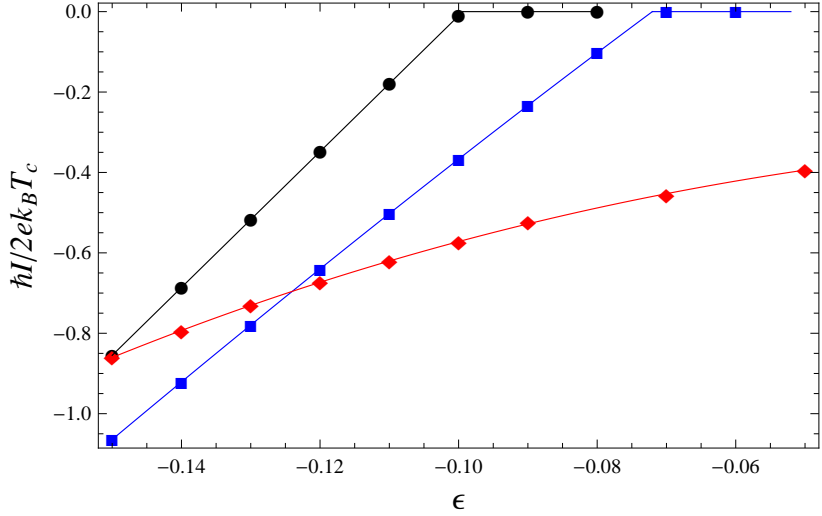}}}
\caption{\label{nofluc}Total current $I$ as a function of the average temperature. The black and the blue curves were obtained ignoring thermal fluctuations and the red curve was obtained taking fluctuations into account. In the case of the black curve the temperature was uniform, whereas for the other curves $\delta=0.2$. The other parameters are as in fig.~\ref{conjec}.}
\end{figure}

We may also inquire how a temperature gradient and/or fluctuations affect the supercurrent. The black and the blue curves in fig.~\ref{nofluc} were evaluated without taking thermal fluctuations into account. In this case the supercurrent becomes the total current, and is independent of position. We note that, for equal average temperatures, a temperature gradient gives rise to larger currents. This result is counterintuitive, since we might expect that the temperature gradient leads to a region where superconductivity is weak (in the case of the blue curve there is a region where $T>T_c$), and that this region limits the supercurrent around the loop. We may attribute our result to proximity, and to the fact that nonuniformity provides greater flexibility to accomodate phase gradients in the order parameter. The red curve shows the total current when fluctuations are properly taken into account. We see that fluctuations reduce the total current for temperatures considerably below $T_c$, but lead to additional current when the temperature increases.

\section{Beyond TDGL\label{BTD}}

\begin{figure}\center{
\scalebox{0.85}{\includegraphics{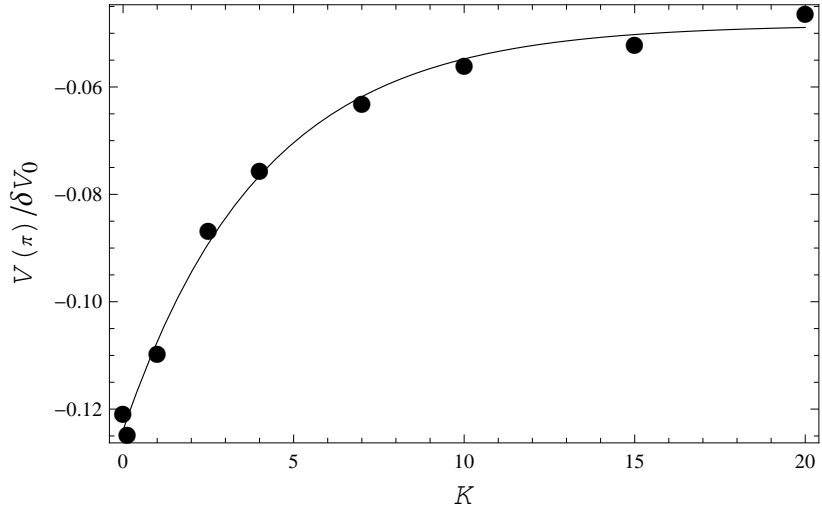}}}
\caption{\label{NKWT} Ratio of voltage to temperature difference, as a function of the Kramer--Watts-Tobin parameter $K$. The dots were calculated and the curve is the empirical fit $V(\pi )/\delta V_0=-0.076\exp(-0.23K)-0.048$. The enclosed flux is $0.3\Phi_0$ and the remaining parameters are as in fig.~\ref{Voffi}.}
\end{figure}

Figure \ref{NKWT} shows the voltage at $\theta =\pi$ as a function of the parameter $K$ defined under eq.~(\ref{Dchi}). We see that $|V(\pi )|$ decreases as $K$ increases, but this trend seems to saturate. Our results are reasonably fitted by the expression $V(\pi )/\delta V_0=-0.076\exp(-0.25K)-0.048$, indicating that the voltage value for large $K$ amounts to about 40\% of its value for $K=0$.

\section{Discussion\label{Dss}}
Our simulations show that when the temperature along a superconducting loop is nonuniform and the loop encloses a magnetic flux that is a noninteger multiple of half the quantum of flux, then a voltage arises in the direction perpendicular to the temperature gradient. We have also investigated the range of parameters over which this effect is most pronounced. For appropriate parameters and for a temperature span of the order of $10^{-1}T_c$, this voltage is of the order of a percent of $V_0=\hbar D/2e\xi_\beta^2$. We have found that this effect is entirely due to thermal fluctuations, and not to the temperature dependence of the Ginzburg--Landau coefficients.

With the estimates in eq.~(\ref{estimates}), we obtain $V_0\sim 2.4\times 10^{-8}(T_c^2/n_e\ell_e w^2)^{1/3}$[volt\,cm$^{2/3}$K$^{-2/3}$]. Taking $n_e\sim 10^{23}{\rm cm}^{-3}$ and reqiring $\xi_\beta =0.5L=3\xi (0)$ leads to $V_0\sim 10^{-5}\, T_c$\,[volt/K]. The conditions $\xi_\beta =0.5L=3\xi (0)$ are difficult to achieve: within the present framework of estimates they give $w\sim 10^{-17}/LT_c\,$[cm$^3$K] and $\ell_e\sim 10^2L^2T_c\,$[cm$^{-1}$K$^{-1}$], leading to $w\sim 10^{-14}\,$cm$^2$ and $\ell_e\sim 10^{-4}\,$cm for $L\sim 10^{-3}\,$cm and $T_c\sim 1\,$K. Requiring instead the less favorable conditions $\xi_\beta =0.2L=7\xi (0)$ would increase $w$ and decrease $\ell_e$ by two orders of magnitude. Larger voltages could be obtained by connecting several rings in series.

As in the case of the field-induced effect (both in conventional and in high $T_c$ superconductors), we have found that the largest Nernst signal is obtained slightly below $T_c$, but persists in a broad range, below and above $T_c$. It is hard to compare between the ordinary Nernst effect and the flux-induced effect, since they are qualitatively different phenomena. In our case the signal depends on the detailed ratios between the sample perimeter and the material characteristic lengths, whereas the field-induced effect can be present in bulk samples. In the the field-induced effect the signal raises with the field until saturation is achieved, or it peaks at a certain field, whereas in our case the signal is an oscillatory function of the flux. It should be noted, though, that for some samples and temperatures in \cite{Wang} there are superimposed oscillations for weak magnetic fields.

Our study differs from those in \cite{HHG,Pb,GZ,ec,Cz} since we consider a uniform material and from that in \cite{Nikul} since we consider a uniform cross section.

\ack
This research was supported by the Israel Science Foundation, grant No.\ 249/10. Numeric evaluations were performed using computer facilities of the Technion---Israel Institute of Technology.

\appendix

\section{Comparison between the field-induced and the flux-induced effects}
It is not obvious how to make a ``fair" comparison between the two effects, because they depend on different quantities. Ideally, we could imagine a situation in which the flux is an appreciable fraction of $\Phi_0$ and the temperature span is an appreciable fraction of $T_c$, but due to the large involved area the magnetic field and the temperature gradient are negligible.

In this appendix we will study the average voltage that appears in a thin square film $0\le x,y\le L'$, $0\le z\le h$, when a uniform field $B=\Phi /L'^2$ is applied in the $z$-direction and a temperature gradient is present in the $y$-direction. In this 2D situation there is no definite circuit as in fig.~\ref{heuristic} that encloses a well defined flux, but due to confinement we may still look for remnants of the flux-induced effect. We will indeed find evidence for situations in which the flux contribution to the Nernst voltage is not negligible in comparison to the expected voltage due to the field.

If the film is sufficiently thin, the induced magnetic field can be neglected and we can take a vector potential that is independent of time; the electric field will be electrostatic and in this appendix will be described by the scalar potential $V$. With the notation of {\it e.g.} \cite{Kopnin}, we can write the free energy in the form
\be
G=h\iint\limits_{\!\!\!\!0\le x,y\le L'}\!\!{\d} x {\d} y \left[ \alpha |\Delta |^2+\frac{\beta }{2}|\Delta |^4+\gamma \hbar^2|(\mathbf{\nabla} -\i {\bf A}')\Delta |^2\right]
\ee
with ${\bf A}'=-[2\pi B(y-L'/2)/\Phi_0]\hat{x}$.

We define a representative length $\xi'_\beta$ and a representative order parameter $\bar\Delta$ for the 2D case in the fluctuation region by requiring $\beta\bar\Delta^4h\xi '^2_\beta =k_BT_c$ and $\beta\bar\Delta^2\xi '^2_\beta  =\gamma \hbar^2$, whence $\xi '_\beta =(\gamma ^2\hbar ^4h/\beta k_BT_c)^{1/2}$ and $\bar\Delta =(k_BT_c/\gamma \hbar^2h)^{1/2}$. Defining a normalized order parameter $\psi =\Delta /\bar\Delta$, the free energy takes the form
\be
G=k_BT_c\iint\limits_{\!\!\!\!0\le x,y\le L'}\!\!{\d} x {\d} y \left[ \frac{T(y)-T_c}{\xi^2(0)T_c} |\psi |^2+\frac{1 }{2\xi '^2_\beta}|\psi |^4+|(\mathbf{\nabla} -\i {\bf A}')\psi |^2\right]\,,
\label{G2D}
\ee
and we shall set $T(y)=[1+\epsilon +\delta (1-2y/L')]T_c$.

In the dirty limit, the Ginzburg--Landau equation reads
\be
(t'_0\partial_t+{\i}V/V'_0)\psi =-\left[ \frac{\xi '^2_\beta (T(y)-T_c)}{\xi^2(0)T_c} +|\psi |^2-\xi '^2_\beta (\mathbf{\nabla} -\i {\bf A}')^2\right]\psi \,,
\label{GLeq}
\ee 
with $t'_0=\xi '^2_\beta /D$ and $V'_0=\hbar /(2et'_0)$. The divergence of the total current vanishes due to electroneutrality and therefore
\be
\nabla^2V/V'_0=5.68\nabla\cdot {\rm Re}[\psi^*(-\i\mathbf{\nabla} - {\bf A}')\psi ] \;.
\label{Poisson}
\ee

We discretize the problem by dividing the square into $N\times N$ cells, such that the center of cell $(i,j)$ is located at $x=(i-1/2)L'/N$, $y=(j-1/2)L'/N$, and denote by $\psi_{i,j}$ and $V_{i,j}$ the values of $\psi$ and $V$ at cell $(i,j)$. For inner cells, $1<i,j<N$, the term $(\mathbf{\nabla} -\i {\bf A}')^2\psi$ in (\ref{GLeq}) can be approximated by
\be
(\mathbf{\nabla} -{\i} {\bf A}')^2\psi \approx (N/L')^2(U_j\psi_{i+1,j}+U^*_j\psi_{i-1,j}+\psi_{i,j+1}+\psi_{i,j-1}-4\psi_{i,j}) \;,
\label{appsq}
\ee
with $U_j=\exp [2\pi{\i} (j-1/2-N/2)\Phi/N^2\Phi_0]$. At the borders, we substitute into (\ref{appsq}) $\psi_{0,j}=U_j\psi_{1,j}$, $\psi_{N+1,j}=U^*_j\psi_{N,j}$, $\psi_{i,0}=\psi_{i,1}$ and $\psi_{i,N+1}=\psi_{i,N}$.

Since for cell $(i,j)$ the discretized value of the right hand side in (\ref{GLeq}) is $-(N^2\xi '^2_\beta/L'^2k_BT_c)(\partial G/\partial \psi^*_{i,j})$, the variance of the Langevin term that adds to either ${\rm Re}[\psi ]$ or ${\rm Im}[\psi ]$ during a lapse of time $\tau$ is 
\be
\langle\eta_{2D}^2\rangle =\frac{N^2\xi '^2_\beta\tau T(y)}{ L'^2T_ct_0} =\frac{N^2D\tau T(y)}{ L'^2T_c}\;.
\label{variance2D}
\ee

We can therefore perform an iteration step for $\psi_{i,j}$ during a lapse of time $\tau$ in three sub-steps: (i) an Euler iteration, in which we add the right hand side of (\ref{GLeq}), multiplied by $\tau /t'_0$; (ii) addition of Langevin terms $\eta_{2D}$ to ${\rm Re}[\psi_{i,j}]$ and to ${\rm Im}[\psi_{i,j}]$; and (iii) approximation of the influence of $V$ by means of the transformation
\be
\psi_{i,j}\rightarrow \frac{1-{\i}eV_{i,j}\tau /\hbar -(eV_{i,j}\tau /\hbar )^2/3}{1+{\i}eV_{i,j}\tau /\hbar -(eV_{i,j}\tau /\hbar )^2/3}\psi_{i,j} \;.
\ee

The right hand side of (\ref{Poisson}) can be discretized as
\be
\rho_{i,j}=5.68(N/L')^2{\rm Im}[\psi_{i,j}^*(U_j\psi_{i+1,j}+U^*_j\psi_{i-1,j}+\psi_{i,j+1}+\psi_{i,j-1})]
\ee
and (\ref{Poisson}) can be solved by successive over-relaxation 
\be
V_{i,j}\rightarrow (1-\omega )V_{i,j}+\frac{\omega }{4}\left(V_{i+1,j}+V_{i-1,j}+V_{i,j+1}+V_{i,j-1}-(L'/N)^2V'_0\rho_{i,j}\right)
\ee
at the inner cells; at the border we have to substitute according to the Neumann condition $V_{0,j}=V_{1,j}$, $V_{N+1,j}=V_{N,j}$, $V_{i,0}=V_{i,1}$ and $V_{i,N+1}=V_{i,N}$. We took values of $\omega$ according to the Chebyshev acceleration sequence. 

Although fluctuations of the electric field are very important in 1D superconductors \cite{Langev}, their influence is usually neglected in higher dimensions. Here they will be ignored  in the zeroth approximation, but will be taken into account as a correction, as explained in the following. 

According to our discussion in section \ref{ress}, if $T(y)$ is replaced with $T(L'/2)$ in eq.~(\ref{variance2D}), the expected time average of $V$ ought to be independent of position. We find, however, small but statistically significant potential differences, which can be attributed to the fact that fluctuations of $V$ were ignored. In order to comply with the second law of thermodynamics, we subtract these potential differences from our results.

\begin{figure}\center{
\scalebox{0.85}{\includegraphics{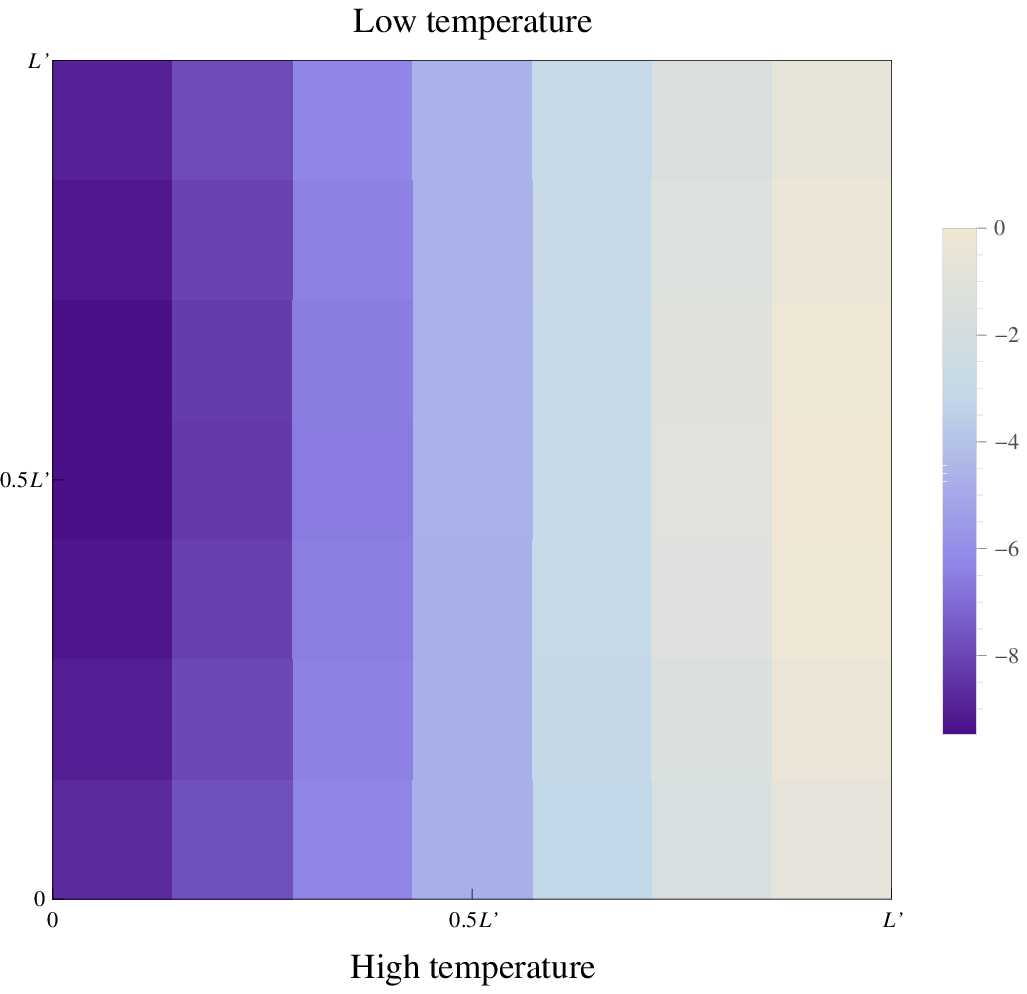}}}
\caption{\label{dens} Density plot of the potential in a thin square sample at average temperature $T_c$, pierced by a magnetic flux $0.7\Phi_0$. $\xi '_\beta =0.5L'=3\xi (0)$; $\delta =0.1$. The color scale is in units of $10^{-3}V'_0$.}
\end{figure}

Figure \ref{dens} is a typical density plot of the potential for a sample in the range that we have studied. As expected, the electric field is essentially perpendicular to the temperature gradient. For the purpose of comparison with our results in the case of a ring, on the following we will focus on the potential difference between the leftmost and the rightmost cells, along the line of average temperature. In the case of a $5\times 5$ partition of the square, these cells are centered at $(x,y)=(0.1L',0.5L')$ and $(x,y)=(0.9L',0.5L')$.

In order to study the field-induced Nernst effect, we consider a very small square, such that the flux is restricted to the range $0<\Phi\leq\Phi_0/8$ and has therefore negligible influence. The inset in fig.~\ref{dens} shows the Nernst electric field $[V(0.1L',0.5L')-V(0.9L',0.5L')]/0.8L'$ divided by the temperature gradient $2\delta T_c/L'$, in units of $V'_0/T_c$, as a function of the magnetic field $B$, in units of $\Phi_0/\xi '^2_\beta$. The geometric and material parameters are not experimentally relevant; they were chosen for illustration purposes. In this region we obtain that the Nernst field is proportional to the magnetic field, and hardly sensitive to the temperature. The Nernst coefficients we obtain are of the order of $10^{-1}D/cT_c$.

The main panel in fig.~\ref{dens} studies samples with the same material parameters and the same magnetic field range as in the inset, but this time the size of the sample was chosen to span a range of flux comparable with $\Phi_0$. In this case we find oscillations in the flux dependence of the Nernst signal, as expected from a quantum flux dependent phenomenon. Unlike the case of the magnetic field influence, these oscillations depend strongly on the average temperature, and may even reverse sign.

\begin{figure}\center{
\scalebox{0.85}{\includegraphics{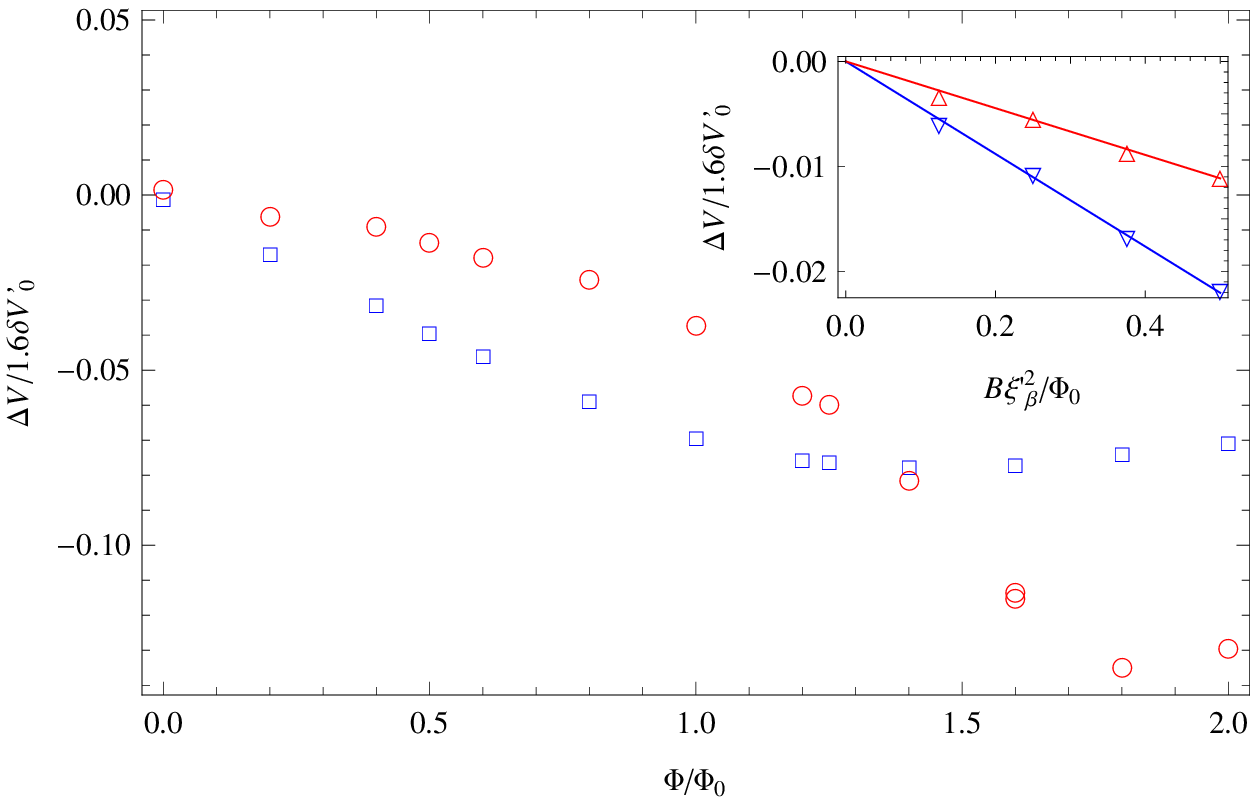}}}
\caption{\label{oscil} Nernst signal as a function of the flux, for different temperatures. $\Delta V$ stands for $V(0.1L',0.5L')-V(0.9L',0.5L')$. The inset shows the Nernst signal as a function of the magnetic field, for different ratios $L'/\xi '_\beta$ (in this case it is more meaningful to regard $\xi '_\beta$ as a fixed parameter). $\square$: $\xi '_\beta=0.5L'$, $\ep =0$; {\large$\circ$\normalsize}: $\xi '_\beta=0.5L'$, $\ep =-0.2$; $\nabla$: $L'=0.5\xi '_\beta$, $\ep =0$ (we also evaluated the Nernst signal for $L'=0.5\xi '_\beta$ and $\ep =-0.2$, and obtained practically the same results); $\triangle$: $L'=0.33\xi '_\beta$, $\ep =-0.2$. In all cases $\xi '_\beta =3\xi (0)$, $\delta =0.1$ and $N=5$. }
\end{figure}

\end{document}